\documentclass{appolb}
\usepackage{graphicx}

\begin{document}
\title{Fission of SHN \& its hindrance: odd nuclei \& isomers.%
\thanks{Presented at the XXXV Mazurian Lakes Conference on Physics, Piaski, Poland, September
3-9, 2017}%
}
\author{W.~Brodzi\'nski, M.~Kowal, J.~Skalski
\address{National Centre for Nuclear Studies, Ho\.za 69,
		PL-00-681 Warsaw, Poland \linebreak}\\ 
P.~Jachimowicz 
 \address{Institute of Physics,
University of Zielona G\'{o}ra, Szafrana 4a, 65516 Zielona
G\'{o}ra, Poland} \\
}


\maketitle
\begin{abstract}
 After shortly analyzing data relevant to fission hindrance of odd-A nuclei and 
 high-$K$ isomers in superheavy (SH) region we point out the inconsistency 
 of current fission theory and propose an approach based on the instanton 
 formalism. A few results of this method, simplified by replacing 
 selfconsistency by elements of the macro-micro model, are given to 
 illustrate its features. 
\end{abstract}
  
\section{Introduction}
 Occurence of isomers - relatively long-lived excited states - 
 is well established in many nuclei, including SH region, see 
  e.g. \cite{HC}. It is believed that the approximate conservation of
  the high - $K$ quantum number (related to the axial symmetry of a nucleus)  
 combined with the low excitation result in the hindrance of their  
 electromagnetic decay. 
 The macro-micro model based on the deformed Woods-Saxon (W-S) potential 
 predicts \cite{XZWW,JKS1,JKS2} high-$j$ orbitals lying close to the Fermi 
 level in $Z=102-110$ nuclei. This explains presence of known isomers and 
 suggests both new ones and high-$K$ ground- or low-lying 
  states in odd and odd-odd nuclei. 
  Such states could live longer than the ground states (g.s.) which makes the 
 study of their stability, and in particular, of their spontaneous fission 
 (SF), very interesting.  

\section{SF hindrance in odd nuclei \& isomers}
   
 It is known that fission half-lives of odd nuclei  
  are 3-5 orders of magnitude longer than those of their even-even 
 neighbours - see e.g. the recent review \cite{FisHess}; 
 slightly smaller odd-even hindrance is observed for fission 
 isomers in actinides \cite{NucTab}. This phenomenon is usually attributed 
 to the specialization energy - increase in fission barrier due to 
 configuration ($K$-number) constraint. Notice, however, that such increase 
 should depend on the $\Omega$ (projection of the ang. momentum on the symmetry
 axis) of the odd orbital, because of smaller level 
 densities for larger $\Omega$, while the data contradict this. 
 \cite{FisHess}. 

  The data on fission hindrance of high-$K$ isomers in heaviest nuclei are 
 given in Table B and Fig. 13 in \cite{Kon}. 
  After eliminating likely erroneous point for $^{262}$Rf - see \cite{FisHess},
  the one for $^{256}$Fm, based on only two observed fission events 
 \cite{256Fm}, and not much informative lower bounds on $T_{sf}(izo)$
 (i.e. much smaller than $T_{sf}(g.s.)$), only data for $^{250}$No \cite{250No} 
 and $^{254}$No \cite{254No} are left.  
 The recent measurement \cite{254Rf} added new data on $^{254}$Rf. 
 All three are given in Tab. 1, and may be prudently summarized by saying 
 that hindrance factors HF$=T_{sf}(izo)/T_{sf}(g.s.)>10$ are possible. 
\begin{table}
\caption{ Fission halflives and hindrance factors for the K-isomers and 
 ground states in the first well.  }
\begin{tabular}{ccccc}
 Nucleus    &  $K^{\pi}$  &  $T_{sf}$(g.s.)  &   $T_{sf}$(izo) & 
  HF$=T_{sf}$(izo)/$T_{sf}$(g.s.)   \\
 \hline
  $^{250}$No \cite{250No} &  $(6^+)$ &  3.7 $\mu$s   &  $> 45  \mu$s & 
 $ > 10$  \\
 \hline
  $^{254}$No \cite{254No} &  $8^-$  &   3$\times 10^4$ s &  1400 s  &
  $\approx \frac{1}{20}$   \\
 \hline
  $^{254}$Rf \cite{254Rf} &  $(8^-)$  &  23 $\mu$s   &  $> 50  \mu$s  &
  $ > 2$   \\
                              &  $(16^+)$  &           &   $> 600  \mu$s  &
  $ > 25 $   \\
\end{tabular}
\end{table}
 Data on multiple fission isomers in even-even actinides \cite{NucTab}, when 
 interpreting higher-lying ones as high-$K$ configurations in the second well, 
 suggest HF$=$1-10 for Pu isotopes and $10^3$-$10^4$ in Cm isotopes. 
 
  \begin{figure*}[th]
 	\begin{center}
 		\includegraphics[scale=0.33]{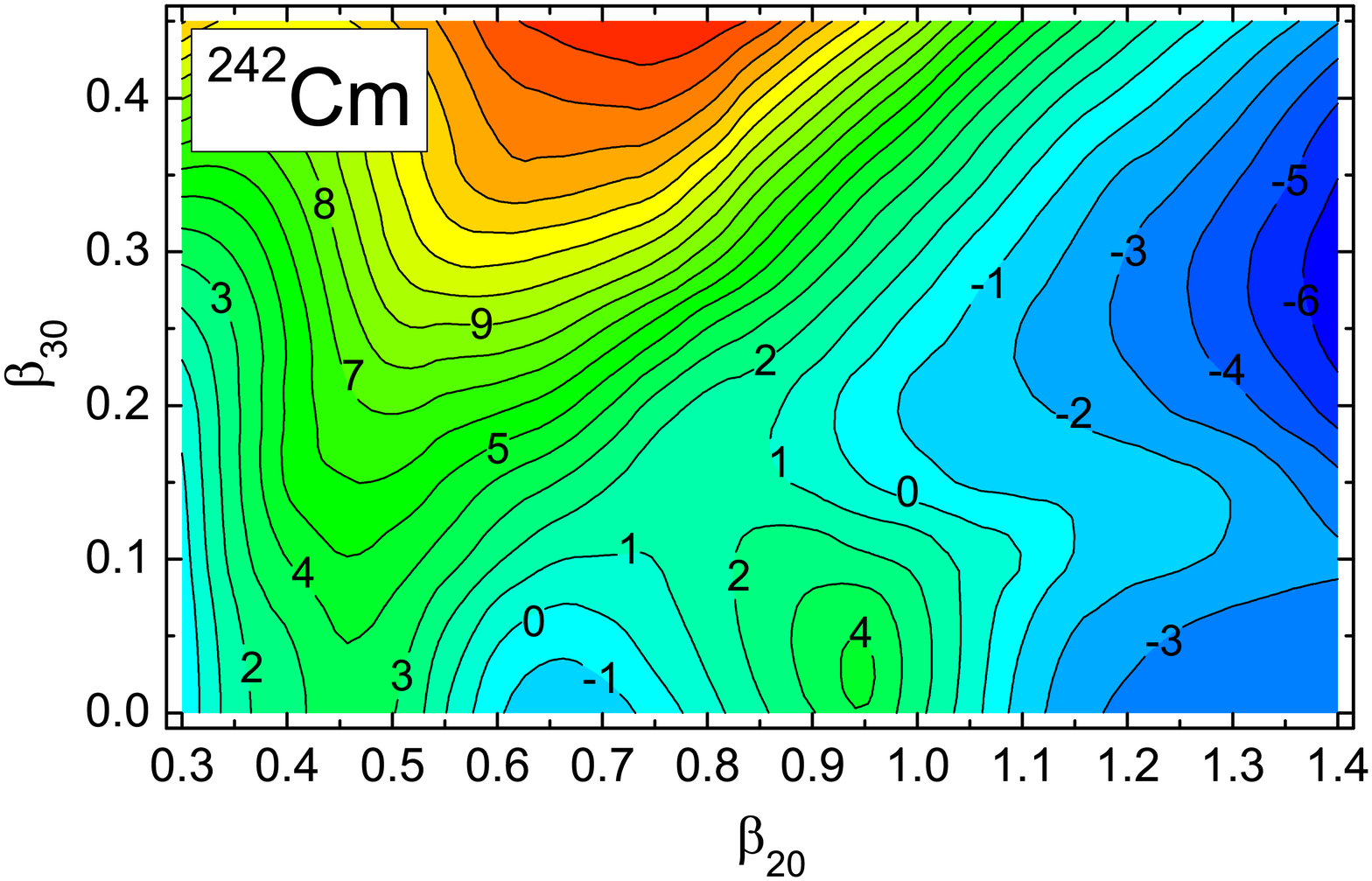}
 		\vspace{-5mm}
 		\includegraphics[scale=0.33]{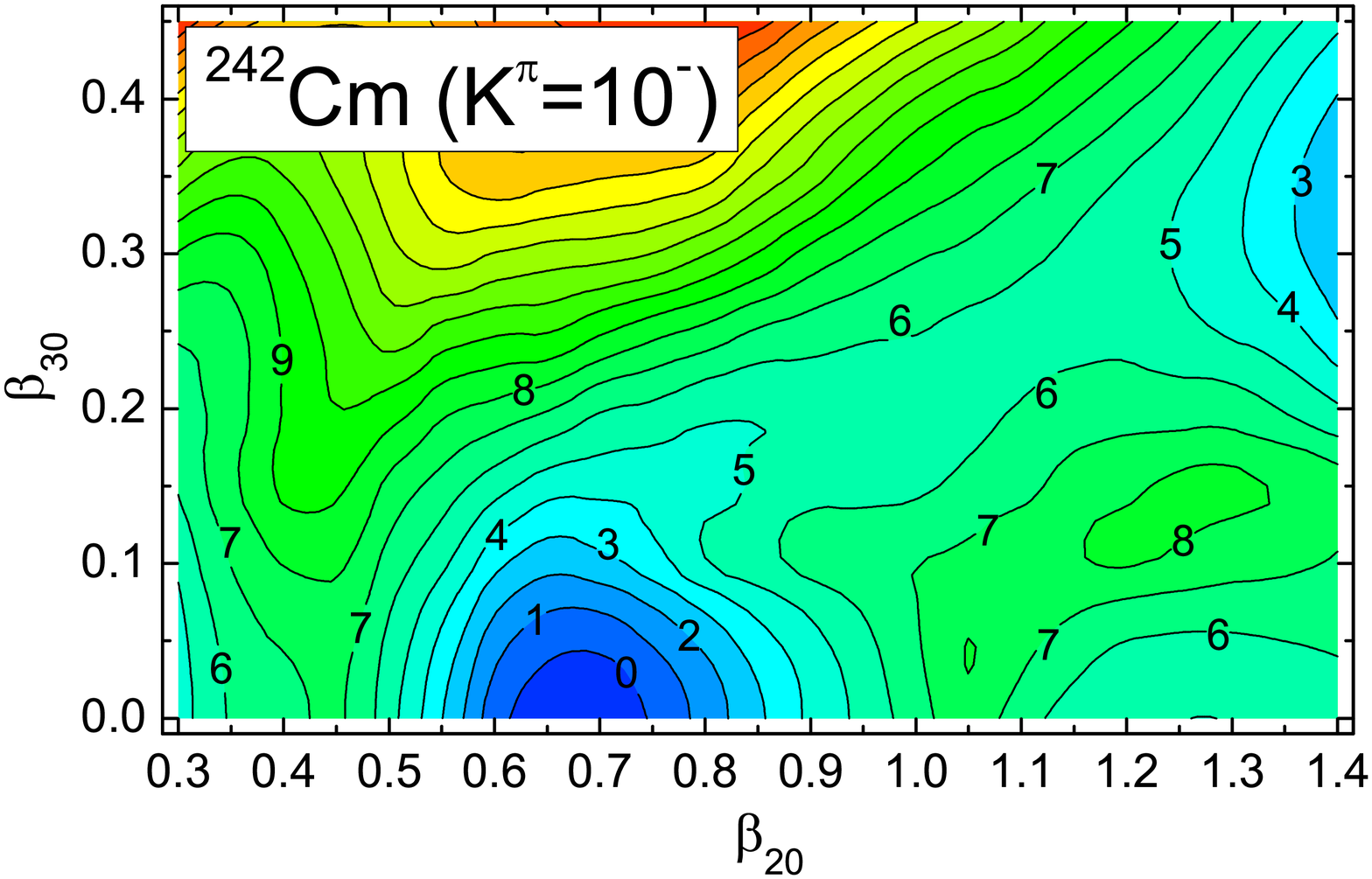} 
 	\end{center}
 \vspace{-8mm} 
 	\caption{ Energy relative to the spherical macroscopic contribution,
 			$E-E_{macr}$(sphere), for the lowest and isomeric $K^{\pi}=10^-$ (parity 
 			at the II-nd minimum) configurations in $^{242}$Cm around and beyond the 
 			second minimum. Seven deformations $\beta_{20}$ - $\beta_{80}$ were included 
 			in the grid; $\beta_{10}$ was fixed by the center of mass condition;   
 			each point results from the minimization over not displayed coordinates. 	}
 	\label{fig:landscp1}
 	
 \end{figure*}

  Within the present theory, the fission hindrance is related to the blocking 
 mechanism: one blocked orbital corresponds to a configuration of an odd 
 nucleus, two blocked orbitals give rise to a 2 quasi-particle isomer in an 
 even-even nucleus, etc. One expects an increase in energy of the isomeric 
 configuration $E_{conf}$, which involves a specialization energy for blocked 
 orbitals, relative to the adiabatic one over a whole region of 
 deformation. In general, this modifies both the shape and height of the isomer 
 fission barrier in comparison to that of the g.s., as it  
  follows from energy landscape $E_{conf}-E_{exc}$, with    
  $E_{exc}$ - the excitation energy of the isomer above the g.s. 
 Specialization energy must depend both on the symmetry of the barrier 
 configuration and the dynamics. While the data suggest that specialization 
  energy increases the barrier in some cases, very large isomeric~vs.~g.s. 
 fission barrier increase is obtained in calculations for many configurations 
 with blocked high-$\Omega$ orbitals. 
 In Fig. 1 we show energy landscapes around and beyond the second minimum in 
 $^{242}$Cm: the adiabatic one and for a fixed $K=10$ state (no intrinsic 
 parity is indicated as the reflection symmetry is broken), corresponding to 
  the $K^{\pi}=10^-$, dominantly $\nu 11/2^+[615],\nu 9/2^-[734]$ configuration 
 in the II-nd well, a unique candidate for a high-$K$ isomer there. 
 Huge rise of the fission barrier height and width for the isomer relative to 
 the adiabatic one can be seen in Fig. 1. In view of this, the 
 experimental relative HF \cite{NucTab} for two shape isomers in $^{242}$Cm 
 might be understood as coming solely from the hindrance of the EM decay to the 
 g.s. in the II-nd well, with the subsequent fission of the latter. Another 
 calculated large rise in barrier due to blocking the high-$\Omega$ 
 orbitals may be seen in Fig. 3 in \cite{JKS2}, this time for the predicted 
 $K^\pi=12^-$ g.s. of the SH odd-odd nucleus $^{272}$Mt.

 Triaxiality of the fission saddle could decrease specialization energy 
 as well as odd-even and isomeric HFs. Another mechanism acting in this 
 direction would be a non-selfconsistent variation of pairing gaps, minimizing 
 the action $\int \sqrt{2B(q)(V(q)-E)}dq$, proposed in \cite{Laz} 
 ($q$ -deformation, $B(q)$ - mass parameter, $V(q)$ - deformation energy and 
 $E$ - g.s. energy). Based on an earlier idea of \cite{MorBa} and calculations 
 \cite{SBPB}, this interesting result is, however, doubtful since:  
 1) the cranking formula for inertia was used as a general one, 2) an analog 
  of the velocity - momentum constraint, crucial for the condition of minimal 
 action, was ignored. As we show below, the lack of a proper inertia  
   parameter is the main obstacle in the treatment of fission of a system 
  with blocked levels.

\section{Failure of the standard SF rate evaluation with blocked states}

 In even-even nuclei, pairing provides an energy 
 gap of at least $2\Delta$ between the g.s. and the lowest 2 q.p. 
 excitation; this amounts to more than 1 MeV in heavy nuclei. 
 One can thus assume that there are no sharp level crossings of a many-body 
 system and that the adiabatic approximation can be applied. This leads to the 
 well-known cranking formula for the inertia parameter, which can be used to 
 compose action integral and minimize it over various fission trajectories.
  
 The situation changes drastically for odd and odd-odd nuclei. In such a case, 
 the neutron or proton contribution to the cranking mass parameter 
 $B_{q_i q_j}$, derived {\it as if the adiabatic approximation were 
  legitimate}, reads:
\begin{eqnarray}
\label{cranking}
B_{q_{i} q_{j}} & = & 2\hbar^{2}
\Bigg[\sum_{\mu,\nu\neq\nu_{0}}
\frac{\langle \mu |\frac{\partial {\hat H}}{\partial q_{i}}|\nu\rangle 
	\langle \nu |\frac{\partial {\hat H}}{\partial q_{j}}|\mu\rangle}
{\left(E_{\mu}+E_{\nu}\right)^3}
\left(u_{\mu}v_{\nu}+u_{\nu}v_{\mu}\right)^2  \\   \nonumber 
& + &\frac{1}{8}\sum_{\nu\neq\nu_{0}}\frac{\left(\tilde{\varepsilon}_{\nu}
	\frac{\partial \Delta}{\partial q_{i}} -\Delta \frac{\partial 
		\tilde{\varepsilon}_{\nu}}{\partial q_{i}}\right) 
	\left(\tilde{\varepsilon}_{\nu}\frac{\partial \Delta}{\partial q_{j}}-
	\Delta \frac{\partial \tilde{\varepsilon}_{\nu}}{\partial q_{j}}\right)}
{E_{\nu}^5}\Bigg]   \\  \nonumber 
& + & 2\hbar^{2} \sum_{\nu\neq\nu_{0}}\frac{\langle \nu|
	\frac{\partial {\hat H}}{\partial q_{i}}|\nu_{0}\rangle \langle \nu_{0}|
	\frac{\partial {\hat H}}{\partial q_{j}}|\nu\rangle}{\left(E_{\nu}-E_{\nu_{0}}
	\right)^{3}}\left(u_{\nu}u_{\nu_{0}}-v_{\nu}v_{\nu_{0}}\right)^{2} . 
\\  \nonumber 
\end{eqnarray}
Here, the ground state corresponds to the odd nucleon 
occupying the orbital $\nu_0$. It is assumed that the one pairing 
gap $\Delta$ and one Fermi energy $\lambda$ describe simultaneously the g.s. 
and its two-quasiparticle excitations: those with the odd particle in the 
state $\nu_0$ (which give contribution in the square bracket) and those with 
the odd particle in the state $\nu\neq\nu_0$ and the orbital $\nu_0$ paired 
(whose contribution is in the third line of the formula). The quantity 
$\tilde{\varepsilon_{\nu}}$ is defined by
$\tilde{\varepsilon_{\nu}}=\varepsilon_{\nu}-\lambda$, $u$ and $v$ are 
the usual BCS occupation amplitudes.
 It is clear that this expression is invalid whenever a close avoided 
crossing is encountered, as the contribution propotional to $(E_{\nu_0}-
E_{\nu})^{-3}$ is nearly singular there. Moreover, due to a partial 
occupation of levels, the singularity may come about from a degeneracy of the 
quasiparticle energies of orbitals at the opposite sides of the Fermi level. 
Already these two reasons make the cranking formula unusable. But there is 
still another deficiency: a departure from the symmetry preserved on a part 
of the fission trajectory produces a negative 
contribution to the inertia parameter whose magnitude would depend on the 
proximity of the relevant level crossing and could dominate the whole expression.  
Therefore a more suitable method which goes beyond the adiabatic approximation is needed. \\

\section{Instanton motivated approach to SF of odd nuclei \& isomers}

Our idea is based on the instanton formalism applied to the SF process, which 
 was formulated for the mean-field setting in \cite{LNP}, \cite{N2} and further
 investigated in \cite{JS}. The instanton equations given there read:
\begin{equation}
\label{equat}
\hbar \frac{\partial \phi_{i}({\tau})}{\partial \tau}=(\zeta_{i}-\hat{h}(\tau))\phi_{i}(\tau) , 
\end{equation} 
which are basically the time dependent Hartree-Fock equations transformed to 
the imaginary time $t\rightarrow -i\tau$ with a periodicity fixing term 
 $\epsilon_{k}\phi_{k}$ (since the bounce solutions should fulfill the 
 periodicity condition $\phi_{k}(-T/2)=\phi_{k}(T/2)$). In these equations,  
 $\phi_{i}$, $i=1,..., N$ are the single-particle (s.p.) states composing the 
 $N$-body Slater state and $\zeta_i$ are the Floquet exponents which for 
 the selfconsistent instanton would be equal to the s.p. energies at the 
 metastable minimum, $\zeta_i=\epsilon_i(q_{min})$. However, for a finite 
 imaginary-time interval $[-T/2,T/2]$, $\zeta_i\ne \epsilon_i(q_{min})$, 
 although they tend to this limit when $T\rightarrow\infty$. 
 The Eq. (\ref{equat}) conserve the overlaps  
 $\langle \phi_i(-\tau)\mid \phi_j(\tau)\rangle=\delta_{i j}$.  
 The instanton action is given by:  
\begin{equation}
\label{sact}
S=\hbar\int_{-T/2}^{T/2}d\tau \sum \limits_{i=1}^{N} \left\langle 
 \phi_{i}(-\tau)\big|\partial_{\tau} \phi_{i}(\tau)\right\rangle = 
 \int_{-T/2}^{T/2}d\tau \sum \limits_{i=1}^{N} \left\langle 
 \phi_{i}(-\tau)\big| \zeta_{i}-\hat{h}(\tau) \big| 
 \phi_{i}(\tau)\right\rangle .
\end{equation}
The Eq. (\ref{equat}) are more difficult to handle than their 
 real-time counterparts since the selfconsistent Hamiltonian 
 $\hat{h}[\phi^{*}(-\tau),\phi(\tau)]$ is now nonlocal in $\tau$. 
 
 Here we replace the selfconsistent mean field in (\ref{equat}) by 
 the phenomenological Hamiltonian with a deformed W-S potential. 
 This can be viewed as a simplification of a selfconsistent theory
 to a macro-micro version. In this approach the collective velocity $\dot{q}$ 
 must be provided as an external information. We take it from: 
 \begin{equation}
 \label{qdot}
 B_{even} (q) {\dot q}^2 = 2 (V(q)- E)  , 
 \end{equation}
 where $q$ is a collective coordinate (e.g. the quadrupole moment) along a 
 chosen path through the barrier, 
 $E$ is the g.s. energy, $V(q)$ - potential energy, 
 and $B_{even} (q)$ - inertia parameter for the neighbouring even-even nucleus.

In solving the equations with the W-S potential we restrict to the 
 subspace of the ${\cal N}$ adiabatic orbitals $\psi_{\mu}(q)$.
 In this subspace, there are ${\cal N}$ bounce
  solutions $\phi_i(\tau)$, each of which tends to the s.p. orbital 
 $\psi_i(q_{min})$ at the metastable minimum as $T\rightarrow \pm \infty$.
 By expanding the solutions onto adiabatic orbitals,  
\begin{equation} 
\label{expan1}
\phi_i(\tau)=\sum_{\mu} C_{\mu i}(\tau) \psi_{\mu}(q(\tau))  ,
\end{equation} 
we obtain the following set of equations for the square matrix of the 
 coefficients $C_{\mu i}(\tau)$: 
\begin{equation}
\label{cequat}
\hbar\frac{\partial C_{\mu i}}{\partial\tau}+{\dot q}
\sum_{\nu} \langle \psi_{\mu}(q(\tau)) \mid\frac{\partial \psi_{\nu}}
{\partial q} (q(\tau))\rangle C_{\nu i} = 
[\zeta_i-\epsilon_{\mu}(q(\tau))] C_{\mu i}  .
\end{equation}
 The conservation of overlaps leads to the condition on $C_{\mu l}(\tau)$: 
\begin{equation} 
\label{over1}
\sum_{\mu=1}^{{\cal N}} C^*_{\mu i}(-\tau) C_{\mu j}(\tau) = \delta_{i j}  .
\end{equation} 
Thus, the quantity $p_{\mu i}(\tau) = C^*_{\mu i}(-\tau)C_{\mu i}(\tau)$ may be
 considered as a quasi-occupation (it can be negative or even complex in 
 general case) of the adiabatic level $\mu$ in the bounce solution $i$, with 
 $\sum_{\mu} p_{\mu i}(\tau) =1$, $\sum_i p_{\mu i} =1$. The action coming 
 from one occupied s.p. bounce state $\phi_i(\tau)$ is:   
\begin{equation}
\label{isact}
S_{i}=\hbar\int_{-T/2}^{T/2}d\tau \sum_{\mu=1}^{\cal N} [\zeta_i-\epsilon_{\mu}
(q(\tau))] p_{\mu i}(\tau) ,
\end{equation}
 and the total action is a sum of the 
 contributions from the occupied s.p. bounce states:
 $S_{tot}=\sum_{i, occ}S_{i}$. 

One can ask whether the instanton action tends to the adiabatic one in the 
 limit of small $\dot{q}$. 
 The comparison of action values for various ${\dot q}$ for a two level system 
 is shown in Tab. \ref{Tab:act2lvl}. 
 The adiabatic action is generally higher than the one obtained from the 
 instanton, but with decreasing ${\dot q}$ both values converge to each other, 
 as one would expect. For stronger interaction between levels (implying smaller
  nonadiabatic coupling) the convergence is even faster.
\begin{table}[t]
	\centerline{
		\begin{tabular}{l|lll|lll}
			$\hbar\dot{q}_{max}/(E_{2}-E_{1})_{min}$ & $0.16$ & $0.08$ & $0.05$ & $0.08$ & $0.04$ & $0.025$ \\
			\hline
			\hline
			$V_{int} [MeV]$ & & $0.5$ &  & &$1.0$ & \\
			\hline
			$S_{inst}/\hbar$ & 1.183 & 0.770 & 0.569 & 0.398 & 0.218 & 0.149  \\
			$S_{adiab}/\hbar$ & 2.015 & 1.007 & 0.672 & 0.459 & 0.229 & 0.152 \\
			\hline
	\end{tabular}}
	\caption{Instanton action values compared with the adiabatic ones in the 2-level system for different maximal velocities $\dot{q}_{max}$ and two values of interaction strength $V_{int}$. Here $(E_{2}-E_{1})_{min}=2V_{int}$ and the ratio $\hbar\dot{q}_{max}/(E_{2}-E_{1})_{min}$ should be sufficiently small for the adiabatic approximation to hold. }
	\label{Tab:act2lvl}
\end{table}

\begin{figure}[htb]
\centerline{%
\includegraphics[angle=-90, width=9.8cm]{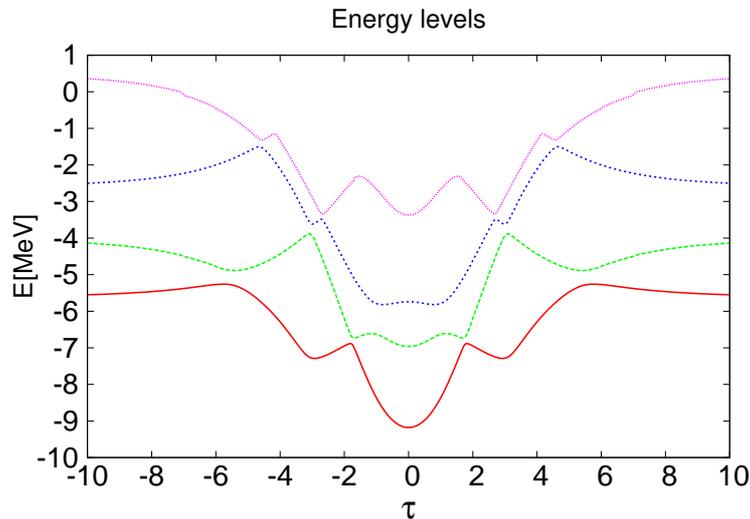}}
\caption{Energies of $1/2^+$ neutron states against the imaginary time 
  determined from ${\dot q}(\tau)$. }
\label{Fig:4lev}
\end{figure}

 We present the behaviour of solutions to Eq. (\ref{cequat}) and resulting 
 action values for four $\Omega^{\pi}=1/2^+$ neutron levels taken from the 
 deformed W-S potential for $^{272}$Mt isotope along the axial 
 (close to static) fission path. The energy levels are depicted in Fig. 
 \ref{Fig:4lev}. The continuous path was determined based on the 
 energy landscape calculated for $\beta_{20}, \beta_{40}$ deformation 
 parameters with the minimization over $\beta_{60}, \beta_{80}$.
\begin{figure}[htb]
	\centerline{%
		\includegraphics[angle=-90, width=9.4cm]{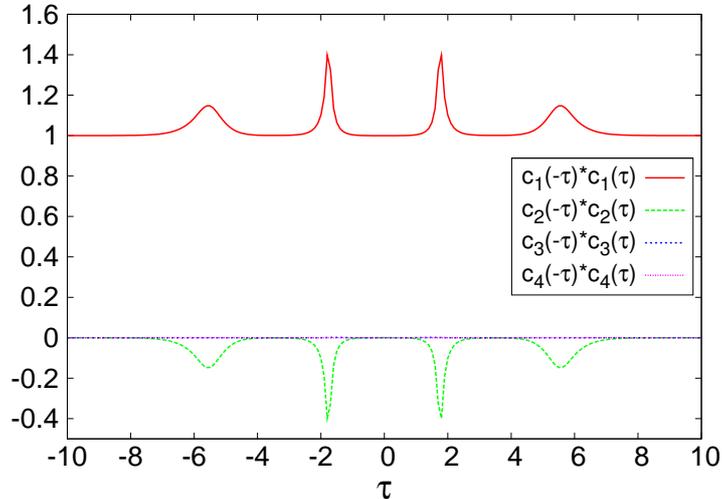}}
	\caption{Quasi-occupations for instanton starting as the lowest 
  adiabatic level.}
	\label{Fig:qocc1}
\end{figure}
The instanton solution starting and ending as the last occupied state below 
 the Fermi level at the minimum (the lowest one in Fig. \ref{Fig:4lev}) 
 is shown in Fig. \ref{Fig:qocc1} in terms of 
 quasi-occupations introduced above. One can see that the particle remains 
 mostly in the 
 initial adiabatic state, except in the vicinity of the avoided crossings 
 where it excites to the second adiabatic level.  
 As long as these crossings are isolated (no other level comes close to them), 
 the excitations to higher states are negligible. This behaviour is in 
 contrast to what we know from the real time dynamics; the closest analogy 
 would be the 2-level Landau-Zener model, where, if the system starts in the 
 lower state at $t=-\infty$ and some nonadiabatic transitions take place 
 during the evolution, then there is a nonzero probability (given by the 
 Landau-Zener formula) that the system will end up in the upper state at       
  $t=+\infty$. \\
\begin{table}[t]	
	\centerline{
	\begin{tabular}{c|lll}
		$\hbar\dot{q}_{max}$ [MeV] & $S_{inst}/\hbar$  & $S_{adiab}/\hbar$  \\
		\hline
		\hline
		0.14 & 2.6818 &  55.048  \\
		0.09 & 2.4892 &  36.699  \\
		0.06 & 2.3492 &  25.689  \\
	\end{tabular}}
\caption{Comparison of the action corresponding to the lowest state obtained from the instanton solution ($S_{inst}$) and in the adiabatic approximation ($S_{adiab}$) for a few values of maximal collective velocity $\dot{q}_{max}$.  }
\label{Tab:act4lvl}
\end{table}
 A comparison of the instanton action (for the above solution) with the 
 adiabatic one is presented in Tab. \ref{Tab:act4lvl} for three different 
 collective velocities $\dot{q}$ - the one from Eq. (\ref{qdot}), and two 
 scaled down by a constant factor. As may be seen, the adiabatic formula 
 overestimates the instanton action, giving the values more than order of 
 magnitude larger. This shows how far from the adiabatic limit we actually 
 are in this case of the unpaired level undergoing sharp avoided crossings.

\begin{figure}[htb]
	\centerline{%
		\includegraphics[angle=-90, width=8cm]{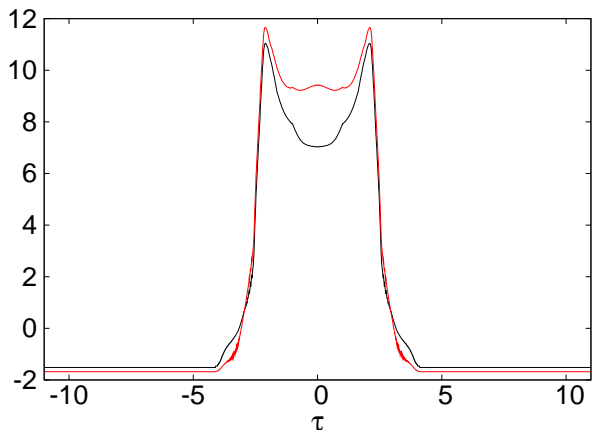}}
	\caption{Comparison of the total action integrands for six (black line)
  and seven (red line) neutrons.}
	\label{Fig:lvlcon}
\end{figure}

 A difference in total action between the odd nucleus and its even-even 
 neighbour comes from: 1) a difference in ${\dot q}$ and 2) the contribution 
 of the last state occupied by the unpaired nucleon. The integrands of 
 the total action for six or seven particles on the lowest four out of 
 ${\cal N}$=8, $\Omega^{\pi}=3/2^{+}$ neutron levels in $^{272}$Mt, with the 
 4-th state empty or singly occupied, are shown in Fig. \ref{Fig:lvlcon}. 
 As one can see, the contribution of the odd nucleon is rather smooth and 
 moderate (in general, it can be negative). Note that contributions to $S$ 
 from other $\Omega$s and parity will still decrease its part in the total.  
  
\section{Conclusions}
 
 Experimental data suggest a mechanism for fission
 hindrance for high-$K$ isomers similar as that for odd-A nuclei 
  in the whole SH region. 
 Pairing plus specialization energy (configuration –
preserving) mechanism seems to have a too strong effect, as judged from 
 energy landscapes for some odd-$A$ nuclei. However, the current description 
 of fission half-lives, employing adiabatic approximation, is not suitable for 
 odd-A nuclei and isomers.   
 The instanton method adapted to the mean-field formalism
 may provide a basis for the minimization of action.
 The preliminary, non-selfconsistent studies indicate that in this method 
 the action is well defined for an arbitrary path and the 
 contribution to action of the odd nucleon is not large. 
 The formalism for paired systems includes dynamic changes of pairing gaps
  as postulated in~\cite{MorBa}, but such that follow from the 
 Hamiltonian-like dynamics \cite{JS}. Their study and work on the inclusion of the selfconsistency are 
 under way.


\end{document}